# Spin-crossover induced ferromagnetism and layer stacking-order change in pressurized 2D antiferromagnet MnPS$_3$

Hanxing Zhang[1,2], Caoping Niu[1,2], Jie Zhang[1,2], Liangjian Zou[1,2], Zhi Zeng[1,2], and Xianlong Wang[*,1,2]

[1] *Key Laboratory of Materials Physics, Institute of Solid State Physics, HFIPS, Chinese Academy of Sciences, Hefei 230031, China.*

[2] *University of Science and Technology of China, Hefei 230026, China.*

\_\_\_\_\_\_\_\_\_\_

*Correspondence to: xlwang@theory.issp.ac.cn

**Abstract:**

High-pressure properties of MnPS$_3$ are investigated by using the hybrid functional, we report a spin-crossover pressure of 35 GPa consisting with experimental observation (30 GPa), less than half of existing report (63 GPa) using the Hubbard *U* correction. Interestingly, a spin-crossover induced antiferromagnetism-to-ferromagnetism transition combined with stacking-order change from monoclinic to rhombohedral are founded, and the ferromagnetism origins from the partially occupied *t$_{2g}$* orbitals. Different from previous understanding, the Mott metal-insulator transition of MnPS$_3$ does not occur simultaneously with the spin-crossover but in pressurized low-spin phase.

# I. Introduction

As typical representatives of 2D van der Waals (vdW) magnetic semiconductors, ferromagnetic chromium halides (CrX$_3$, X = Cl ~ I) and antiferromagnetic transition-metal phosphorous trichalcogenides (TMPX$_3$, TM = Mn ~ Ni and X = S or Se) attract wide attention [1-10] due to potential applications in spintronics [11] and valleytronics [12]. Tuning the magnetic properties of CrX$_3$ and TMPX$_3$ are essential to expand their applications and to explore the magnetic physics in 2D lattice. As a clean and maneuverable technique, pressurization was proven to be an effective way to tune the properties of 2D magnets and widen their applications. For example, at 1.7 GPa, the interlayer antiferromagnetic coupling between two CrI$_3$ layers are switched to ferromagnetic interaction combined with monoclinic-to-rhombohedral stacking-order transition [13]. Furthermore, the properties of TMPX$_3$, magnetic Mott insulators, were also proven to be sensitive to the external pressure. High-pressure measurements show that in (Mn,Fe)P(S,Se)$_3$, transition from high-spin (HS) to low-spin (LS) state, named as spin-crossover, will give rise to 10~20% drastic volume reduction, metal-insulator transition (MIT), and superconductivity [3,7,14]. The abundant property variations indicate that TMPX$_3$ are good models for illustrating the physics in 2D magnetic Mott insulators. Note that, the switch of intralayer magnetic coupling from anti-ferromagnetism (AFM) to ferromagnetism (FM) is not realized yet in TMPX$_3$, FM coupling is highly expected and may notably expand their applications.

Theoretical investigations based on the first-principle methods have focused on TMPX$_3$ properties heavily [14-17]. Nevertheless, the mean-field type exchange-correlation functionals (e.g., GGA) can not sufficiently describe the electronic structures of strongly correlated materials because of the underestimated state localization in GGA calculations. Hubbard $U$ correction (e.g., GGA+$U$) is widely used to increase the $d$ states localization of transition metal compounds [18]. However, in some cases, the GGA+$U$ calculations could not correctly predict the spin-crossover behavior of magnetic strongly correlated materials. For example, the simulated spin-crossover pressure of MnPS$_3$ based on the GGA+$U$ is 63 GPa [17], which is much larger than experimental observation (30 GPa) [7]. The main reason may be that as an empirical parameter, the $U$ value is sensitive to the crystal field and spin state [19].

Since the orbital occupations and state localizations are different, the LS state should have different effective $U$ value with the HS state.

In this work, we use the hybrid functional to study the property evolutions of pressurized $MnPS_3$. The reasons for using hybrid functional are presented as following: First, Hatree-Fock method overestimates the localization of states, and the hybrid functional admixed with a portion of Hartree–Fock exact exchange in GGA can give reasonable state localization. Second, in the spin-crossover process of $Mn^{2+}$ at the octahedral crystal field, two $d$ electrons in $e_g$ orbital will overcome the bandgap and transfer to the spin-down $t_{2g}$ orbitals (Fig. 1). We believe that regardless of the spin-crossover driven forces (crystal splitting or bandwidth broadening) [19], correct determination of the band gap at any pressure is the key to successfully predict the spin-crossover pressure. We can imagine that in the limit, the spin-crossover can't occur if the band gap is infinite. Since the sate localization is sufficiently described, the hybrid functional can give more accurate band gap than the GGA [20].

## II. Methods

All calculations are performed based on the density functional theory with the projector-augmented plane wave (PAW) potentials [21] as implemented in the Vienna Ab-initio Simulation Package (VASP). Heyd-Scuseria-Ernzerhof functional (HSE06) [22] incorporating the separation parameter of 0.2 and the default mixing parameter of AEXX=0.25 is employed. The vdW-D2 correction [23] is included to improve the description of vdW interactions. The energy cutoff for the plane-wave basis-sets and total energy convergence is set to 400 eV and 1×10$^{-5}$ eV, respectively, and a 4×4×4 k-mesh is used for sampling the first Brillouin zone. Based on the HSE06 functional, all investigated structures are fully relaxed until the forces acting on each atom are less than 0.005 eV/Å, and the total stress tensor is reduced to 0.01 GPa.

## III. Results and discussion

Based on the HSE06 functional with AEXX=0.25, we find that comparing with FM, Stripe-type AFM and Zigzag-type AFM states, Néel-type AFM state with a magnetic moment of 4.46 $\mu_B$ is the ground magnetic state of MnPS$_3$ at 0 GPa consisting with neutron scattering measurement [24]. Furthermore, as shown in Fig. 2, the predicted band gap (2.95 eV) agrees well with experimental value (3.0 eV) [25]. The lattice constants are also correctly predicted (See Table I) [26-28]. The results indicate that HSE06 functional can sufficiently describe the MnPS$_3$ properties at ambient condition.

The partial density of states (PDOS) of AFM MnPS$_3$ at 0 GPa is shown in Fig. 3 (a). We can find that Mn's $d$ orbitals are half-filled, and three (two) of five $d$ electrons occupy spin-up $t_{2g}$ ($e_g$) orbitals, while all spin-down $d$ orbitals locate above the Fermi level. Depending on the Goodenough-Kanamori-Anderson (GKA) super-exchange rule [29-31], the AFM state should be the ground state of the insulating MnPS$_3$ with half-filled $d$ orbitals. Furthermore, because the $e_g$ orbitals extend along Mn-S bonds, the spin-up $e_g$ orbitals and S's 3$p$ orbitals contribute equally to the hybridized states just below the Fermi level. The unoccupied states closed to the Fermi level are mainly from the spin-down $t_{2g}$ orbitals indicating that MnPS$_3$ is a Mott insulator with strong Mn-S hybridization.

To investigate the MnPS$_3$ property evolution under high-pressure, the GGA+$U$ framework with a typical 4 eV $U$ value was applied for Mn's 3$d$ states in existing work [17]. However, as shown in Fig. 4, the calculated spin-crossover pressure based on the GGA+$U$ ($U$ = 4 eV) is 63 GPa, which is more than twice as much as experimental observation (30 GPa) [7]. Note that, as shown in Fig. 2, a very large $U$ value of ~11 eV is needed to reproduce the experimentally measured band gap, and 4 eV $U$ correction is not enough. The spin-crossover pressure of MnPS$_3$ will increase with the applied $U$ values increasing because of the band gap enlargement, e.g., the calculated spin-crossover pressure based on the GGA+$U$ ($U$ = 4 eV) is 63 GPa, which is ~53 GPa higher than the counterpart of GGA (Fig. 4). Therefore, a larger spin-crossover pressure than 63 GPa can be expected if ~11 eV $U$ correction is applied.

In the framework of HSE06 (Fig. 4), the MnPS$_3$ maintains HS state with a magnetic moment of ~4.19 $\mu_B$ up to 35 GPa, and it spontaneously change to the LS state (~1.06 $\mu_B$) at 40 GPa resulting in a spin-crossover pressure of ~35 GPa consisting well with experimental observation (30 GPa) [7]. The spin-crossover causes a 12.4% volume reduction comparable to experimental report of 19.8% volume reduction, and the volume reduction origins mainly from the decrease of interlayer distance [Fig. 5 (a)]. Furthermore, interlayer slipping, which occur simultaneously with the spin-crossover, causes a phase transition from monoclinic stacking-order [*C2/m* symmetry, see Fig. 3 (c)] to rhombohedral stacking-order [*R-3* symmetry, see Fig. 3 (d)]. Differing from the monoclinic MnPS$_3$, the unit-cell of rhombohedral MnPS$_3$ contains three staggered layers. The MnPS$_3$ in the LS state, which is named as LS MnPS$_3$ hereafter, at high pressure has the same structure with MnPSe$_3$ at ambient condition (rhombohedral phase with *R-3* symmetry) [32].

The pressure and temperature induced monoclinic-to-rhombohedral stacking-order transition had been widely observed in 2D materials due to the variation of stacking-order [8,13,33,34], for example, pressurization can give rise to a monoclinic-to-rhombohedral phase transition in CrI$_3$ bi-layers combined with an interlayer AFM-to FM transition [13]. However, calculations using the GGA+*U* did not find the spontaneous monoclinic-to-rhombohedral stacking-order transition in MnPS$_3$ [17], and HSE06 is more efficient than GGA+*U* for searching the stable phase of MnPS$_3$ at high-pressure. The possible reason is that the localization of all orbitals in MnPS$_3$ are sufficiently described based on the HSE06, however, the GGA+*U* framework only increases the localization of Mn's *d* orbitals.

Interestingly, togethering with the spin-crossover and monoclinic-to-rhombohedral stacking-order transition, an intralayer magnetic coupling transition [Fig. 5 (b)] from AFM to FM is observed. There is no imaginary frequency in the phonon dispersions of LS-FM MnPS$_3$ with rhombohedral stacking-order at 40 GPa, and the new phase is kinetic stable [Fig. 5 (c)]. The monoclinic-to-rhombohedral stacking-order transition accompanying the interlayer AFM-to-FM transition was observed in CrI$_3$ [8,13,35,36] indicating that stacking-order and magnetic coupling in 2D magnets are associated. The enthalpy difference between FM and AFM state ($\Delta_{FM-AFM}$, negative and positive value correspond to FM and AFM coupling, respectively) is −35 meV/Mn at 40 GPa comparable to that of CrI$_3$ (−26.5 meV/Cr) and CrGeTe$_3$ (−39 meV/Cr) [37]. As shown in the PDOS of LS-FM MnPS$_3$ at 40 GPa [Fig. 3 (b)], spin-down $t_{2g}$ orbitals in the LS state are partially occupied while $e_g$ orbitals are unoccupied. From GKA super-exchange rule [29-31], ferromagnetic coupling can be expected if the $d$ orbitals are partially occupied. With the pressure increasing, the distance between Mn ions become shorter, resulting in stronger direct interaction and higher AFM stability. Therefore, the $\Delta_{FM-AFM}$ increases linearly with pressure increasing from 40 GPa to 60 GPa, and the AFM becomes the ground state again at 60 GPa. Note that, the structural properties are not sensitive to the magnetic coupling, and the LS MnPS$_3$ has rhombohedral stacking-order no matter in FM or AFM state. Following, we will discuss the evolution of band gap and micro-structure.

The band-gap is shown as a function of pressure in Fig. 6 (a), and pressurization smoothly decreases the band-gap of HS-AFM MnPS$_3$ from 2.95 eV at 0 GPa to 0.99 eV at 35 GPa. After transferring to the LS state, a band gap of 0.57 eV is observed in the LS-FM MnPS$_3$ at 40 GPa [Fig. 6 (c)], where all three Mn-Mn distances are almost equal [Fig. 6 (b)]. The spin-crossover causes a semiconductor-to-semiconductor transition but not previous reported MIT [7,17]. The MIT was believed to occur simultaneously with the spin-crossover togethering with a Mn-Mn dimerization in previous simulation [17], where magnetic (AFM-to-FM) and phase (monoclinic-to-rhombohedral) transition was not founded. With pressure increasing to 50 GPa, Mn-Mn dimerization is observed in LS-FM MnPS$_3$, and one Mn-Mn distance is about 0.2 Å shorter than the other two [Fig. 6 (b)]. Furthermore, due to the strong direct interaction among Mn ions in the dimerized Mn-Mn pair, band-gap close is realized in the LS-FM MnPS$_3$ at 50 GPa [Fig. 6 (a) and 6 (d)]. These results indicate that Mn-Mn dimerization and Mott MIT are sensitive to the magnetic state and structure.

## IV. Conclusions

Our results illustrate a success of the hybrid functional in predicting the spin-crossover pressure of MnPS$_3$ magnetic Mott insulator, and spin-crossover induced simultaneous intralayer AFM-to-FM magnetic transition and monoclinic-to-rhombohedral stacking-order transition is found. The Mn-Mn dimerization and Mott metal-insulator transition occur in the pressurized LS MnPS$_3$ but not simultaneously with the spin-crossover. MnPS$_3$ and its family are good models for illustrating the cooperation of structural, electronic, magnetic properties in the Mott insulator. The hybrid functional is an efficient method to study the spin-crossover behaviors of magnetic Mott insulator.

**Acknowledgements:**

This research was supported by the NSFC under Grant of 11674329, Science Challenge Project No. TZ2016001. The calculations were performed in Center for Computational

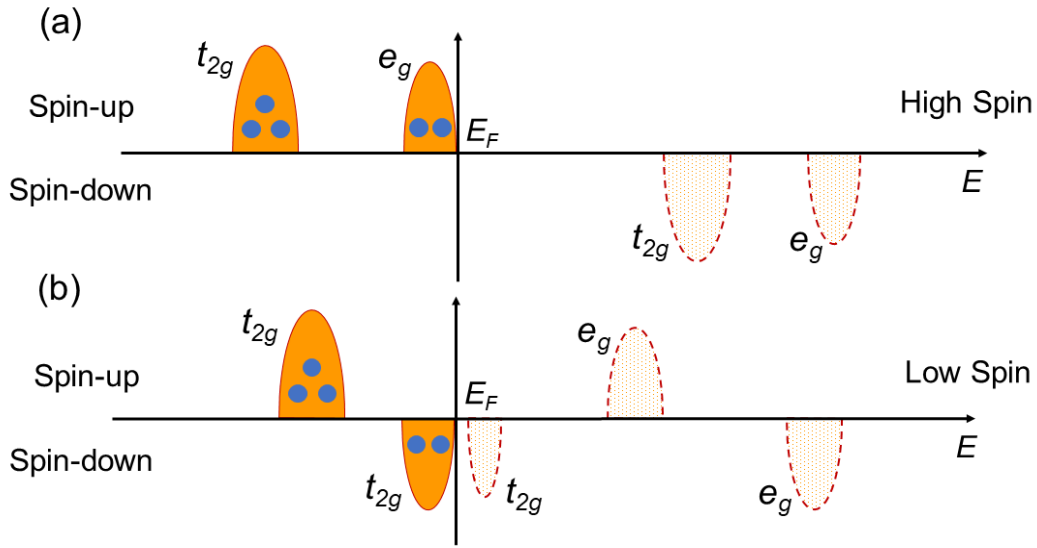

**Fig. 1. (a)** For the high-spin $Mn^{2+}$ at octahedral crystal field, spin-up $t_{2g}$ and spin-up $e_g$ orbitals are occupied by five electrons, and all spin-down orbitals locate above the Fermi-level. **(b)** For the low-spin $Mn^{2+}$ at octahedral crystal field, three and two electrons occupy the spin-up and spin-down $t_{2g}$ orbitals, respectively. In the spin-crossover process, two electrons at the spin-up $e_g$ orbital will overcome the band gap and transfer to the spin-down $t_{2g}$ orbitals.

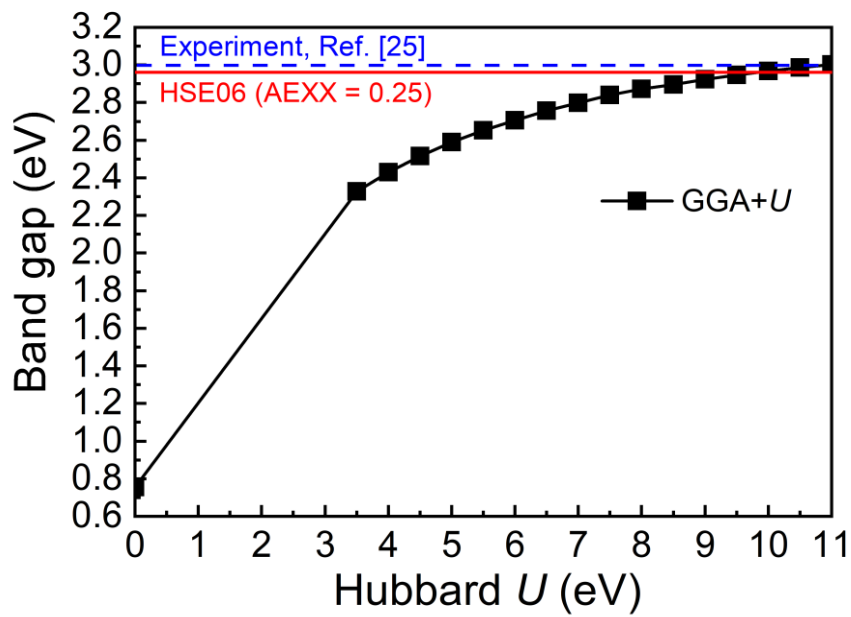

**Fig. 2.** Calculated band gaps by using the GGA+$U$ method are shown as a function of the Hubbard $U$. The horizontal blue dash line represents the experimentally measured band gap (3 eV) at ambient condition (Ref. [25]). The horizontal red solid line shows the simulated band gap (2.95 eV) by using the HSE06 at 0 GPa, which agree well with experimentally measured values. However, a 11 eV $U$ value is needed to reproduce experimental value.

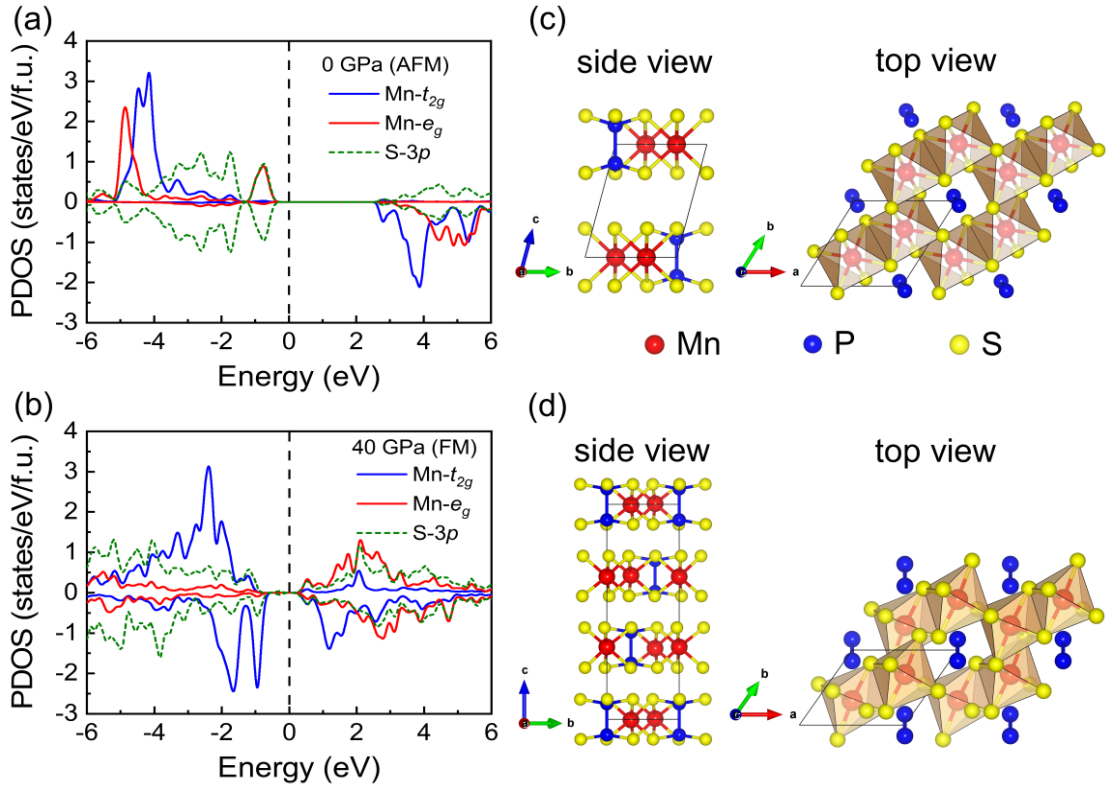

**Fig. 3. (a)** Partial density of states (PDOS) of antiferromagnetic MnPS$_3$ in the high-spin state with monoclinic stacking-order at 0 GPa. **(b)** PDOS of ferromagnetic MnPS$_3$ in the low-spin state with rhombohedral stacking-order at 40 GPa. **(c)** The side and top views of antiferromagnetic MnPS$_3$ with monoclinic stacking-order at 0 GPa. **(d)** The side and top views of ferromagnetic MnPS$_3$ with rhombohedral stacking-order at 40 GPa. In **(a)** and **(b)**, vertical dashed lines present the Fermi level. In **(c)** and **(d)**, balls in red, blue, and yellow colors represent manganese, phosphorus, and sulfur atoms, respectively.

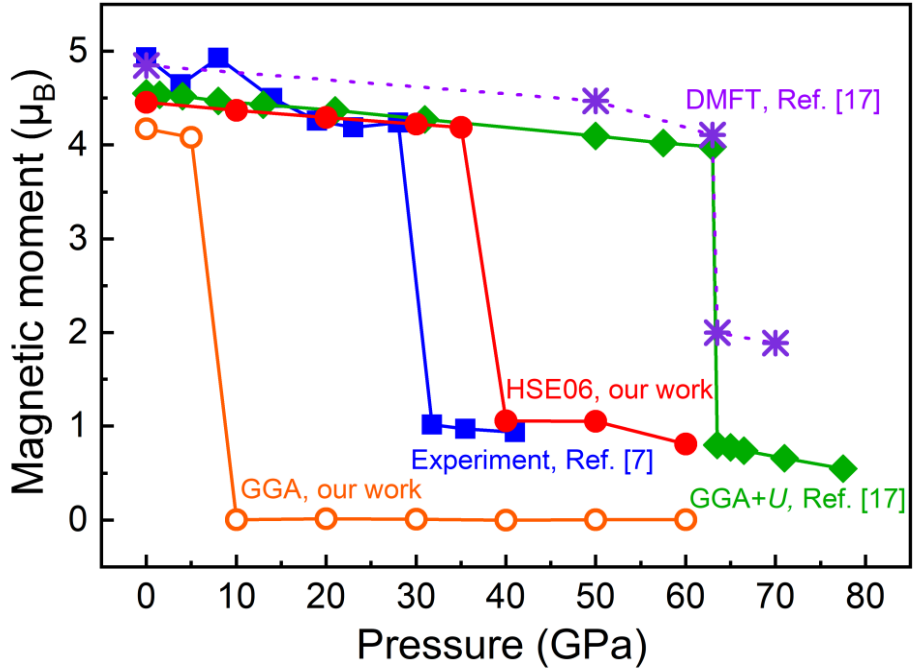

**Fig. 4.** The experimentally measured and theoretically simulated magnetic moments of Mn are shown as a function of pressure. The experimental data is shown as blue squares, the existing theoretical results derived from DMFT and GGA+$U$ ($U$ = 4 eV) calculations are shown in purple crosses and green diamonds, respectively. Our results based on the HSE06 (GGA) are presented in red solid circles (orange hollow circles). The calculated spin-crossover pressure based on HSE06 is ~35 GPa consisting with experimental observations (30 GPa), which is significantly smaller than that from DMFT or GGA+$U$ calculations (63 GPa). The GGA(PBE) calculations gives the smallest spin-crossover pressure of 10 GPa.

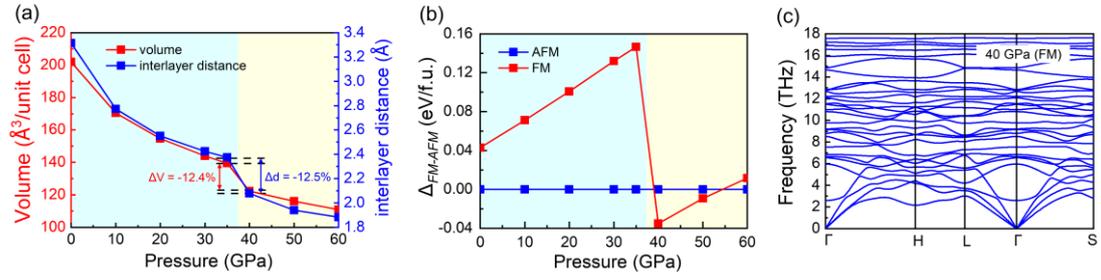

**Fig. 5. (a)** Volume (red, left Y-axis) and interlayer distance (blue, right Y-axis) are presented as a function of pressure. **(b)** Change of enthalpy difference between FM and AFM state ($\Delta_{FM-AFM}$) is shown. **(c)** Phonon dispersions of low-spin ferromagnetic MnPS$_3$ with rhombohedral stacking-order at 40 GPa, and all frequencies are positive. In **(a)** and **(b)**, blue (yellow) region indicates the stability range of high-spin MnPS$_3$ with monoclinic stacking-order (low-spin MnPS$_3$ with rhombohedral stacking-order). The spin-crossover induced volume reduction is mainly caused by interlayer distance decrease, and a simultaneous magnetic transition from antiferromagnetic to ferromagnetic can be founded.

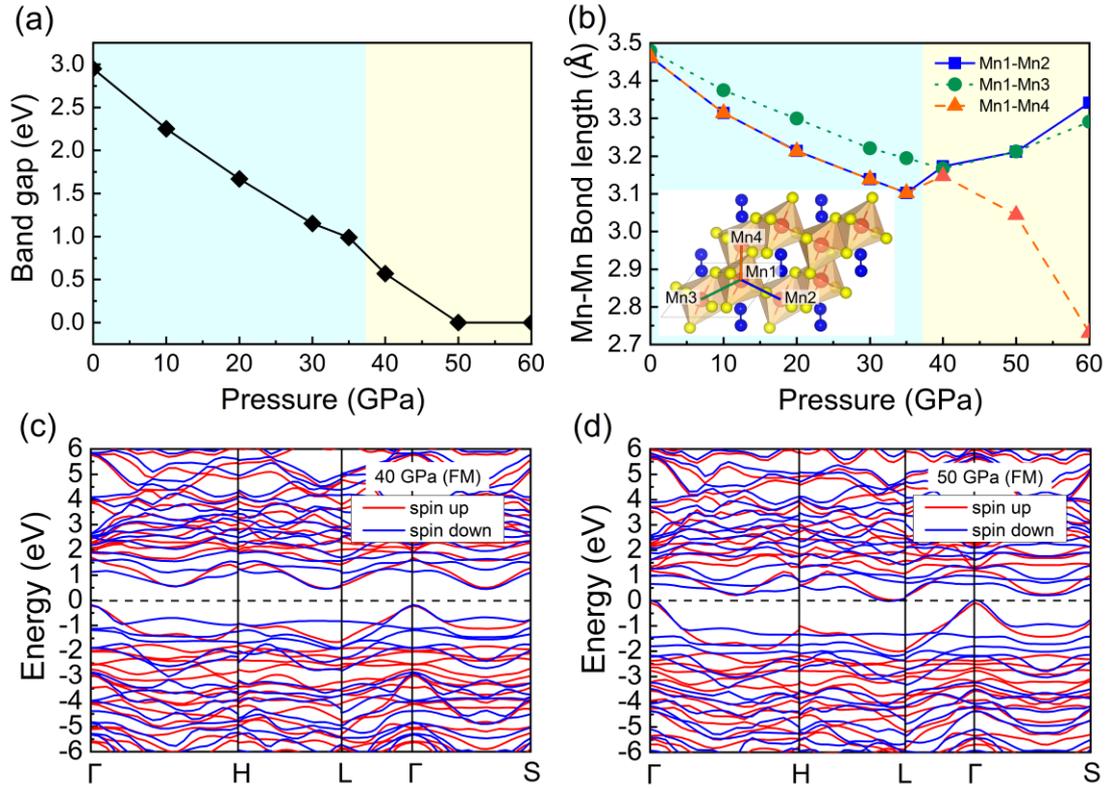

**Fig. 6. (a)** Change of band gap with pressure is shown. **(b)** Mn-Mn bond lengths are presented as a function of pressure, and the top view of the low-spin ferromagnetic $MnPS_3$ with Mn-Mn dimerization at 50 GPa is shown in the inset. **(c)** The band structure of low-spin ferromagnetic $MnPS_3$ with rhombohedral stacking-order at 40 GPa. **(d)** The band structure of low-spin ferromagnetic $MnPS_3$ with rhombohedral stacking-order at 50 GPa. In **(a)** and **(b)**, blue (yellow) region indicates the stability range of high-spin $MnPS_3$ with monoclinic stacking-order (low-spin $MnPS_3$ with rhombohedral stacking-order). The metal-insulator transition and Mn-Mn dimerization occurs in the pressurized low-spin $MnPS_3$ at 50 GPa.

Table Ⅰ. Crystal structures parameters of high-spin MnPS$_3$ with *C2/m* symmetry (monoclinic stacking) at 0 GPa and low-spin MnPS$_3$ with *R-3* symmetry (rhombohedral stacking) at 40 GPa.

| Pressure (GPa) | Lattice Parameters (Å, °) | | | | Atoms | Atomic coordinates (Fractional) | | |
|---|---|---|---|---|---|---|---|---|
| 0 C2/m | a = 6.008 | 6.077[a] | 6.051[b] | 6.009[c] | Mn (4g) | 0 | 0.3328 | 0 |
| | b = 10.406 | 10.524[a] | 10.523[b] | 10.591[c] | P (4i) | 0.0560 | 0 | 0.1693 |
| | c = 6.760 | 6.796[a] | 6.802[b] | 6.805[c] | S (8j) | 0.2417 | 0.1693 | 0.2522 |
| | β = 106.97 | 107.35[a] | 107.32[b] | 107.27[c] | S (4i) | 0.7583 | 0 | 0.2505 |
| | α = γ = 90 | | | | | | | |
| 40 R-3 | a = b = 5.475 | | | | Mn (6c) | 0.6667 | 0.3333 | 0.9968 |
| | c = 14.136 | | | | P (6c) | 0.3333 | 0.6667 | 0.7392 |
| | α = β = 90 | | | | S (18f) | 0.9729 | 0.3012 | 0.7598 |
| | γ = 120 | | | | | | | |

[a] Ref. [26]
[b] Ref. [27]
[c] Ref. [28]